\documentclass[sigconf]{acmart}
\usepackage{booktabs} % For formal tables
\usepackage{amsmath}
\usepackage{amsfonts}
\usepackage{comment}
\usepackage{footmisc}
\usepackage{mathrsfs}
\usepackage{graphicx}
\usepackage{subfigure}
\usepackage{multirow}
\usepackage{algorithm}
\usepackage[most]{tcolorbox}
\usepackage{algorithmic}
\usepackage{multicol}  
\usepackage{enumitem}
\usepackage{array}
\usepackage{float}
\usepackage{hyperref}
\usepackage{xspace}
\tcbuselibrary{breakable}   % 配合文本加框，使得框可以跨页
\newtcolorbox{AIbox}[1]{
  enhanced,
  colback=black!5, % Box background color (light gray)
  colframe=black!50, % Box frame color (darker gray)
  fonttitle=\bfseries, % Title font is bold
  title=#1, % The title of the box
  boxsep=5pt,
  left=5pt,
  right=5pt,
  top=5pt,
  bottom=5pt,
  arc=2mm % Rounded corners
}

\newcommand{\kc}[1]{\textcolor{black}{#1}}

\usepackage{color, xspace}

\AtBeginDocument{%
	\providecommand\BibTeX{{%
			\normalfont B\kern-0.5em{\scshape i\kern-0.25em b}\kern-0.8em\TeX}}}

\copyrightyear{2026}
\acmYear{2026}
\setcopyright{cc}
\setcctype{by-nc-nd}
\acmConference[WWW '26]{Proceedings of the ACM Web Conference 2026}{April 13--17, 2026}{Dubai, United Arab Emirates}
\acmBooktitle{Proceedings of the ACM Web Conference 2026 (WWW '26), April 13--17, 2026, Dubai, United Arab Emirates}
\acmPrice{}
\acmDOI{10.1145/3774904.3792235}
\acmISBN{979-8-4007-2307-0/2026/04}

\begin{document}
\title{To Search or Not to Search: Aligning the Decision Boundary of Deep Search Agents via Causal Intervention}

%% --- 作者 1: Wenlin Zhang ---
\author{Wenlin Zhang}
\orcid{0000-0003-1809-8264}
\affiliation{%
  \institution{City University of Hong Kong}
  \city{Hong Kong}
  \country{China}
}
\email{wl.z@my.cityu.edu.hk}

\author{Kuicai Dong}
\orcid{0000-0002-5564-0641}
\affiliation{%
  \institution{Huawei Technologies Ltd.}
  \city{Shenzhen}
  \country{China}
}
\email{dong.kuicai@huawei.com}

\author{Junyi Li}
\orcid{0009-0007-0480-5593}
\affiliation{%
  \institution{City University of Hong Kong}
  \city{Hong Kong}
  \country{China}
}
\email{junyili@cityu.edu.hk}

\author{Yingyi Zhang}
\orcid{0000-0001-9062-3428}
\affiliation{%
  \institution{City University of Hong Kong}
  \city{Hong Kong}
  \country{China}
}
\email{yzhang6375-c@my.cityu.edu.hk}

%% --- 作者 6: Xiaopeng Li ---
\author{Xiaopeng Li}
\orcid{0009-0008-6162-8500}
\authornote{Corresponding authors.}
\affiliation{%
  \institution{City University of Hong Kong}
  \city{Hong Kong}
  \country{China}
}
\email{xiaopli2-c@my.cityu.edu.hk}

%% --- 作者 5: Pengyue Jia ---
\author{Pengyue Jia}
\orcid{0000-0003-4712-3676}
\affiliation{%
  \institution{City University of Hong Kong}
  \city{Hong Kong}
  \country{China}
}
\email{jia.pengyue@my.cityu.edu.hk}

\author{Yi Wen}
\orcid{0000-0002-5924-1429}
\affiliation{%
  \institution{City University of Hong Kong}
  \city{Hong Kong}
  \country{China}
}
\email{yiwen23-c@my.cityu.edu.hk}

\author{Derong Xu}
\orcid{0000-0002-3971-9907}
\affiliation{%
  \institution{City University of Hong Kong}
  \city{Hong Kong}
  \country{China}
}
\email{derongxu2-c@my.cityu.edu.hk}

%% --- 作者 8: Maolin Wang ---
\author{Maolin Wang}
\orcid{0000-0002-0073-0172}
\affiliation{%
  \institution{City University of Hong Kong}
  \city{Hong Kong}
  \country{China}
}
\email{morin.wang@my.cityu.edu.hk}

%% --- 作者 9: Yichao Wang ---
\author{Yichao Wang}
\orcid{0000-0001-7053-8269}
\authornotemark[1]
\affiliation{%
  \institution{Huawei Technologies Ltd.}
  \city{Shenzhen}
  \country{China}
}
\email{wangyichao5@huawei.com}

%% --- 作者 11: Ruiming Tang ---
\author{Yong Liu}
\orcid{0000-0001-9031-9696}
\affiliation{%
  \institution{Huawei Technologies Ltd.}
  \city{Shenzhen}
  \country{China}
}
\email{liu.yong6@huawei.com}

%% --- 作者 12: Xiangyu Zhao (通讯作者) ---
\author{Xiangyu Zhao}
\authornotemark[1]
\orcid{0000-0003-2926-4416}
\affiliation{%
  \institution{City University of Hong Kong}
  \city{Hong Kong}
  \country{China}
}
\email{xianzhao@cityu.edu.hk}

\renewcommand{\shortauthors}{Wenlin Zhang et al.}

\begin{abstract}
Deep search agents, which autonomously iterate through multi-turn web-based reasoning, represent a promising paradigm for complex information-seeking tasks. However, current agents suffer from critical inefficiency: they conduct excessive 
\kc{searches as they cannot accurately judge}
% search rounds due to poorly calibrated judgments of 
when to stop searching and \kc{start answering}. This stems from outcome-centric training that prioritize final results \kc{over the search process itself.}
% while neglecting intermediate decision quality. 
We identify the root cause as misaligned \textbf{decision boundaries}, the threshold determining when accumulated information \kc{ suffices to answer.}
% becomes sufficient to answer queries. 
\kc{This causes \textbf{\textit{over-search}} (redundant searching despite sufficient knowledge) and \textbf{\textit{under-search}} (premature termination yielding incorrect answers).}
% This misalignment manifests in two failure modes: \textbf{over-search}, where agents redundantly search despite sufficient knowledge, and \textbf{under-search}, where they terminate search prematurely and provide incorrect answers. 
To address these errors, we propose a comprehensive framework comprising two key components. First, we introduce \textbf{causal intervention-based diagnosis} that identifies boundary errors by comparing factual \kc{and counterfactual trajectories}
% trajectories against counterfactual alternatives 
at each decision point. Second, we develop \textbf{D}ecision Boundary \textbf{A}lignment for Deep \textbf{S}earch agents (\textbf{DAS}), which constructs preference datasets from causal feedback and aligns policies via preference optimization. Experiments on public datasets demonstrate that decision boundary errors are pervasive across state-of-the-art agents. Our DAS method effectively calibrates these boundaries, \kc{mitigating} both over-search and under-search to achieve substantial gains in accuracy and efficiency. Our code and data are publicly available at: \url{https://github.com/Applied-Machine-Learning-Lab/WWW2026_DAS}.

\end{abstract}

% \begin{abstract}
 % Our code is available
% at: \footnote{\label{foot1}\url{https://anonymous.4open.science/r/MMAgent-D8E2/}}

% \textbf{Relevance Statement:} This paper proposes \name, a multimodal user agent that aims to overcome the limitations of current user simulators for real-world A/B testing in interactive recommendation scenarios.
% To address the current lack of datasets that simulate authentic recommendation environments, we constructed a multimodal movie recommendation dataset and developed a corresponding recommendation user interface.

% \end{abstract}




\maketitle
\section{Introduction}
\label{sec:intro}
The integration of Large Language Models (LLMs) with autonomous agents has revolutionized intelligent information systems. 
Beyond text generation, agent-based paradigms are now reconstructing behavior spaces~\cite{shi2024prompt, wang2025behavior, zhang2025notellm, zhang2026exploring}, optimizing sequential decisions~\cite{zhao2019deep, zhao2018deep, zhao2018recommendations, fu2023unified, liu2023exploration}, enhancing ranking tasks~\cite{li2023agent4ranking}, and advancing personalization~\cite{li2025survey}.
Focusing on complex information seeking, Agentic RAG is emerging as a promising paradigm~\cite{huang2025deep, li2025towards}.
Unlike traditional RAG systems, it enables sophisticated interactions with search tools~\cite{xu2024enhancing}, including adaptive search invocation, iterative query decomposition and refinement~\cite{diao2025temporal}, systematic validation of retrieved content, and resolution of conflicts between parametric and web knowledge~\cite{huang2025deep, zhang2024agentic, li2025search}. While such retrieval frameworks can also incorporate knowledge graphs~\cite{xu2024multi,yang2022gammae} and multimodal information~\cite{dong-mmdocir, dong2025mmdocrag} to enhance RAG performance~\cite{xu2025harnessing}, the critical advancement lies in the autonomous search capability. Recent reasoning LLMs have begun to internalize these search capabilities~\cite{guo2025deepseek, yang2025qwen3}, enabling the development of \textit{deep search agents}~\cite{jin2025your, jin2025search, jiang2025deepretrieval, dong2025docresearcher} that support complex reasoning and in-depth research workflows.
\begin{figure}[!t]
    \centering
    \includegraphics[width=0.9\columnwidth]{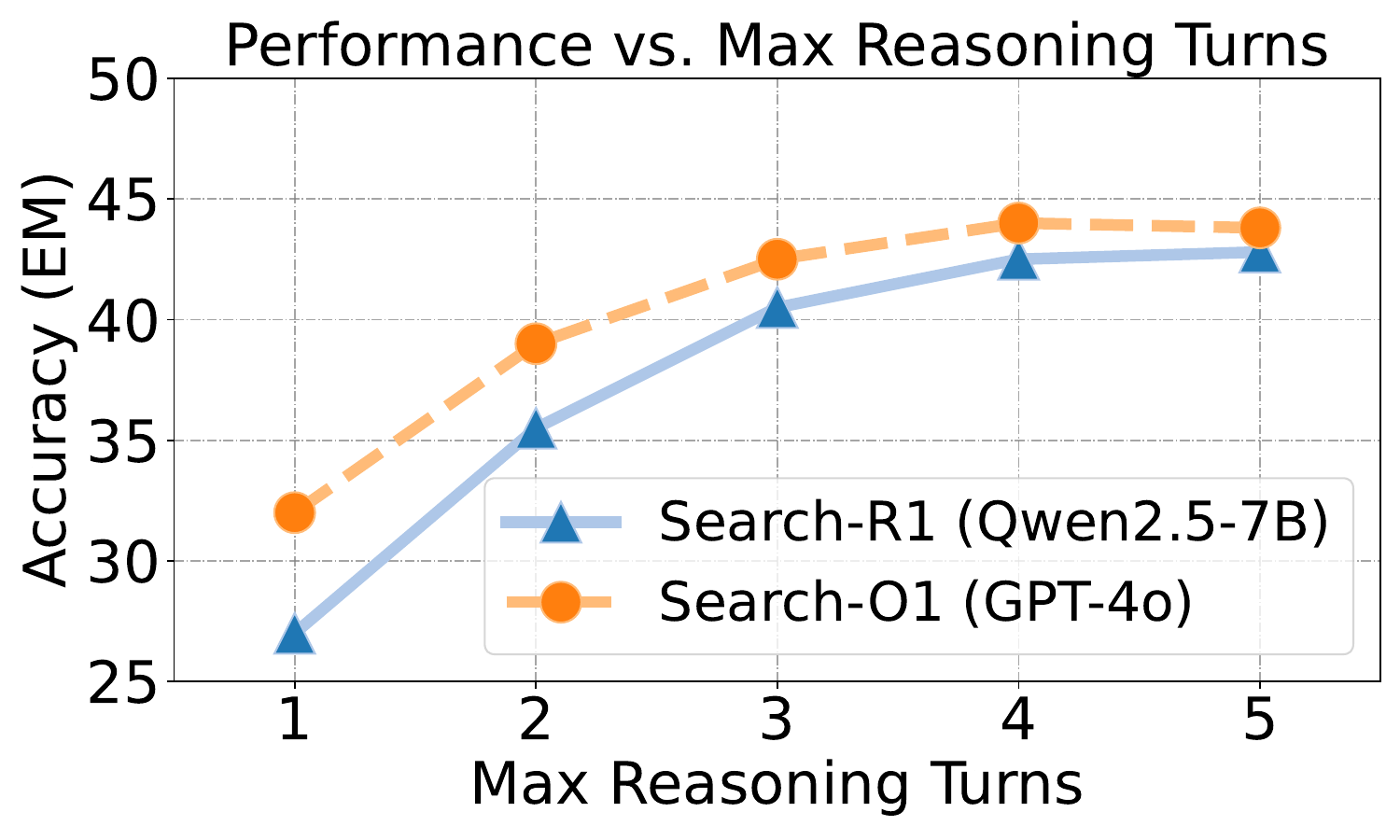}
    % \vspace{-1em}
    \caption{Trade-off between search depth and accuracy on HotpotQA, illustrating the diminishing gains.}
    \label{fig:tradeoff_plot}
\end{figure}

Despite their promise, current deep search agents lack efficiency, often conducting excessive search rounds to solve tasks. This inefficiency stems from outcome-centric training and evaluation paradigms that prioritize final results over intermediate decision quality~\cite{jin2025your, jin2025search, sun2025zerosearch, wang2025stepsearch}.
By treating the reasoning process as a black box, these approaches incentivize deep search agents to search repeatedly as long as the task is eventually solved, without considering the decision optimality of individual round: \textit{whether to \kc{continue searching} or to answer}. However, our empirical study reveals that search gains diminish as rounds increase, \kc{which is especially significant after 3 rounds} as shown in Figure~\ref{fig:tradeoff_plot}. 
While accuracy improves with more reasoning turns, the marginal benefit of each additional search decreases substantially.
This finding underscores a critical challenge: determining the optimal decision boundary that identifies when to stop searching.

We formally define the \textbf{\textit{search decision boundary ($\S$\ref{sec:formal_definition})}} as the critical threshold determining when 
\kc{parametric knowledge and accumulated retrieved information,}
% accumulated information, combining internal parametric knowledge and retrieved external evidence, 
becomes sufficient to answer the query. 
This boundary is pivotal for balancing the accuracy-efficiency trade-off: while additional search rounds may improve accuracy, they incur higher latency and computational costs. However, current deep search agents struggle to optimize this boundary, leading to two prevalent failure modes. 
\textbf{Over-search} occurs when agents conduct redundant searches despite possessing sufficient information, driven by over-reliance on external tools and neglecting internal knowledge~\cite{wang2025otc}. Methods like WebResearcher~\cite{qiao2025webresearcher}, WebSailor~\cite{li2025websailor}, and DeepDive~\cite{lu2025deepdive} achieve performance gains through extensive tool use but suffer from this ``\textit{cognitive offloading}''~\cite{risko2016cognitive}, failing to prevent unnecessary searches. 
Conversely, \textbf{under-search} occurs when agents prematurely terminate search and provide incorrect answers. Knowledge boundary-based approaches like FLARE~\cite{jiang2023active} and Adaptive RAG~\cite{jeong2024adaptive} attempt to prioritize internal knowledge through confidence estimation and routing strategies.
\kc{While they can successfully reduce the usage of search tools, they tend to be over-confident on internal knowledge,}
% , yet they inadequately internalize external search capabilities, 
leading to premature search termination. 
Neither paradigm effectively aligns the search decision boundary to optimize both accuracy and efficiency.

To address these decision boundary errors, we propose a comprehensive framework that systematically diagnoses and corrects suboptimal search decisions \kc{through two key components.}
% Our approach is built on two key components. 
First, we introduce \textbf{causal intervention-based diagnostic methodology ($\S$\ref{sec:probes})} that identifies both over-search and under-search errors by comparing an agent's factual trajectory against counterfactual alternatives. Specifically, when an agent searches, we \kc{we simulate the outcome had it answered instead;}
% intervene to simulate what would have happened had it answered instead; 
when it answers incorrectly, we simulate the outcome of continued search. By evaluating these counterfactual outcomes, we can retrospectively determine whether each decision was causally optimal given the agent's latent knowledge state. 
Second, building upon this diagnostic foundation, we develop the \textbf{Decision Boundary Alignment (DAS) ($\S$\ref{sec:dpo_construction})}.
\kc{DAS} constructs targeted preference pairs from the causal feedback (via pairing suboptimal factual trajectories with their superior counterfactual counterparts) and aligns agent's policy via preference optimization. This alignment process directly teaches agent to internalize when its knowledge is sufficient to answer versus when \kc{further} search is \kc{needed}. 
Extensive experiments on three public datasets demonstrate that DAS \kc{can} successfully calibrate the decision boundary, reducing both \kc{over-search and under-search errors while significantly improving} accuracy and efficiency. In summary, our contributions are in threefold:
\begin{itemize}[leftmargin=*, itemsep=0pt, parsep=0pt, topsep=0pt]
    \item \textbf{Problem Formulation and Analysis.} We formally define and characterize the two primary decision boundary errors, over-search and under-search, and demonstrate through extensive experiments that they constitute a fundamental bottleneck in current deep search agents.
    
    \item \textbf{Causal Diagnostic Methodology.} We propose a novel diagnostic approach using \textit{causal intervention} to rigorously identify both error types by comparing factual and counterfactual trajectories at each decision point.
    
    \item \textbf{Decision Boundary Alignment (DAS).} We introduce DAS, which leverages causal feedback to construct preference datasets and optimize searching policies, achieving substantial gains in both accuracy and efficiency.
\end{itemize}

\section{Decision Boundary Detection and Alignment}
\label{sec:perception}
\begin{figure*}[!htbp]
    \centering
    \includegraphics[width=0.98\linewidth]{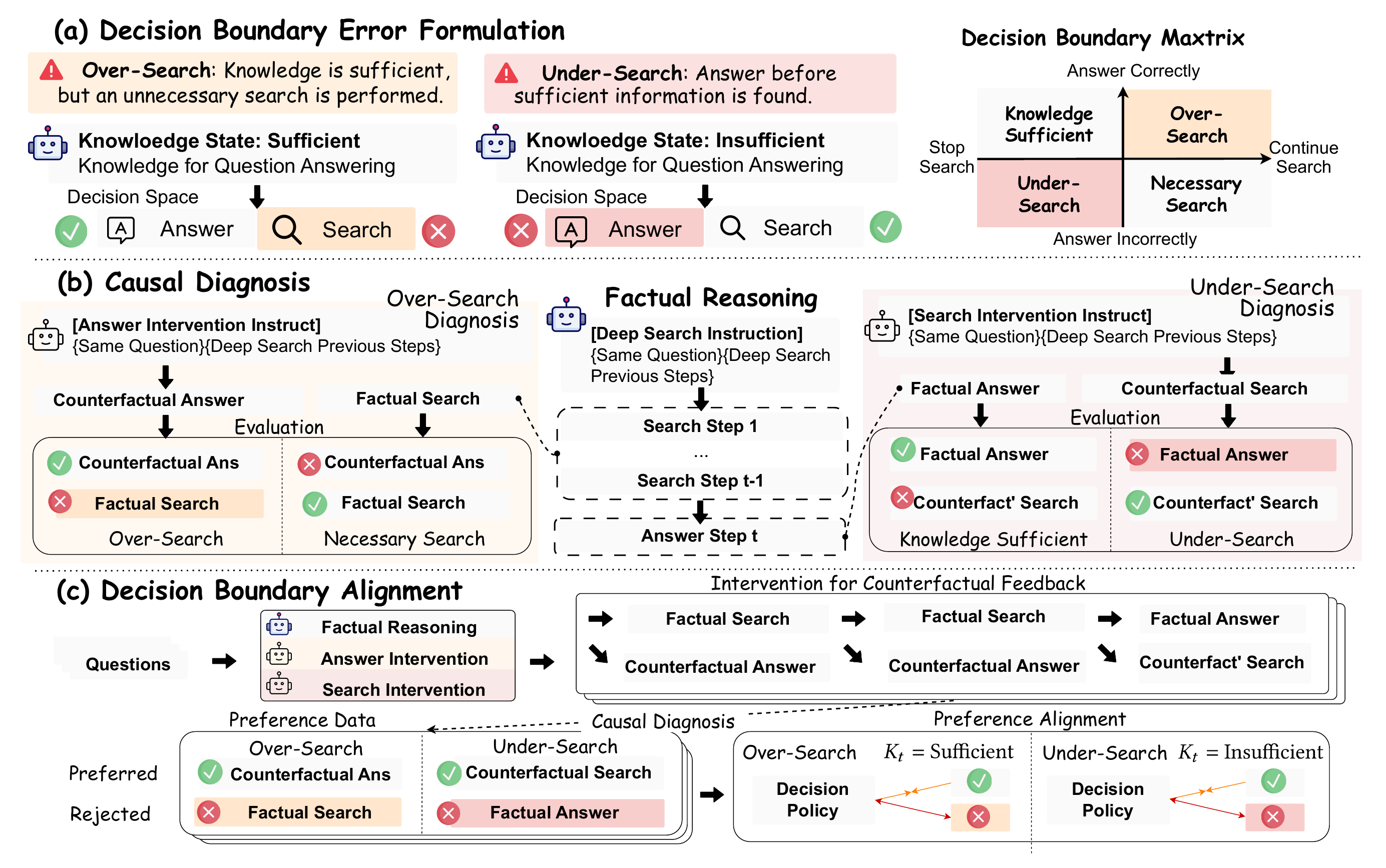}
    
    \caption{The overall framework of DAS. (a) Formal definition of over-search and under-search as the two primary failure modes at the agent's decision boundary. (b) Causal Intervention for decision errors diagnosis within an agent's trajectory through the use of counterfactual instruction. (c) Decision Boundary Alignment for Search Agent based on a preference dataset from identified errors.}
    \label{fig:framework}
\end{figure*}

% （a）拼成一个decision space
% （a）answer-> Not Search（Continue Search Stop Search）
% (a) 符号一致（使用对错）
% (a) 加一个混淆矩阵
% （b）体现干预过程（反事实，因果描述对齐）
% （c）数据构造与体现对齐，压缩前面空间，提升c部分空间

This section presents our approach to align the agent's decision-making with optimal decision preference. We formalize the decision boundary and define two key \textbf{decision errors}: \textbf{over-search} and \textbf{under-search} ($\S$\ref{sec:formal_definition}). To mitigate these errors, we then introduce our method for decision boundary detection ($\S$\ref{sec:probes}) and our Decision Boundary Alignment for Search (DAS) algorithm ($\S$\ref{sec:dpo_construction}). Figure~\ref{fig:framework} illustrates the overall framework.

\subsection{Formal Definition of the Decision Boundary}
\label{sec:formal_definition}

To rigorously assess the decision-making capabilities of a deep search agent, we move beyond holistic trajectory evaluations and adopt a causal lens focused on step-wise optimality. We model the decision process at step $t$ as an interaction between the agent's chosen action $A_t \in \{\texttt{Search}, \texttt{Answer}\}$ and the unobservable ground truth of its knowledge state $K_t \in \{\text{Sufficient}, \text{Insufficient}\}$. As conceptually visualized in Figure~\ref{fig:framework}(a), the decision boundary is not merely a theoretical line but the precise threshold where the agent must distinguish between the necessity of further exploration and the sufficiency of current information. We ground these quadrants in a formal utility framework to mathematically define the conditions that render specific actions suboptimal.

\paragraph{\textbf{Utility and Error Formulation}}
The agent's objective is to maximize the expected utility $U(A_t, K_t)$, which is governed by a set of rewards and costs. We posit that a correct answer yields a high positive reward $R_{\text{correct}}$, whereas providing an incorrect answer incurs a significant penalty $R_{\text{incorrect}}$. Conversely, the act of searching is nuanced: it incurs a computational and operational cost $C_{\text{search}}$, but offers a positive information value $R_{\text{info}}$ when the current knowledge is lacking. This formulation allows us to categorize decision failures not just as wrong outcomes, but as specific deviations from optimal utility:

\begin{itemize}[leftmargin=*, itemsep=4pt, parsep=0pt, topsep=4pt]
    \item \textbf{Over-Search (Efficiency Failure):} An over-search error is strictly defined by the state where the agent chooses to \texttt{Search} despite $K_t$ already being \text{Sufficient}. In this scenario, the search action is mathematically inferior because the information gain is zero, yet the cost $C_{\text{search}}$ is still incurred. The alternative action, \texttt{Answer}, would have immediately yielded the maximum reward $R_{\text{correct}}$ without further expense. Thus, over-search represents a failure of \textbf{efficiency}, where the agent sacrifices immediate operational success for unnecessary computational effort \textbf{that provides no marginal benefit to the answer's quality}.
    
    \item \textbf{Under-Search (Accuracy Failure):} Conversely, an under-search error occurs when the agent chooses to \texttt{Answer} while $K_t$ remains \text{Insufficient}. This decision is catastrophic, as the premature termination of the search process triggers the penalty $R_{\text{incorrect}}$. Here, the alternative \texttt{Search} action is the dominant strategy because the value of acquiring missing information ($R_{\text{info}}$) logically outweighs the search cost ($C_{\text{search}}$). By failing to recognize the need for further exploration, the agent compromises the factual validity of its response. This represents a fundamental \textbf{failure of accuracy} driven by a flawed assumption of sufficiency, ignoring that the prospective payoff from searching is rationally superior.
\end{itemize}

This causal framework transforms the decision-making process into an optimization problem. The agent's goal is to align its policy $\pi(S_t)$ with the latent state $K_t$, ensuring it selects the action that mathematically dominates the alternative in each scenario.

\subsection{Causal Diagnosis of Decision Boundary Errors}
\label{sec:probes}

% \zxy{what are the Challenges/Motivations of the section (e.g., why the problem is important, why it is challenging, what are the shortcomings of existing solutions, why this module is needed, why do you use the provided technology, etc), otherwise your technical contributions look trivial.} 
Localizing decision boundary errors is challenging since unselected actions cannot be observed.
To diagnose the decision errors defined by the unobservable latent state $K_t$, we employs causal intervention, a principled method from causality theory for estimating the counterfactual outcomes of alternative actions~\cite{pearl2009causality}. The diagnostic process is retrospective. Each decision point $t$ is revisited after an agent's trajectory is complete to apply an intervention, which is formalized by the \texttt{do()}-operator as $do(A_t := a')$. The resulting counterfactual outcome is then compared against the factual one to infer the value of $K_t$ and identify the error. The causal diagnostic process is now detailed for each of the two error types: over-search and under-search.

\paragraph{\textbf{Diagnosing Over-search}}
An over-search error is diagnosed at a state $S_t$ when the agent's factual action is $a_t = \texttt{Search}$. The diagnosis hinges on determining if the latent state was actually $K_t = \text{Sufficient}$. To resolve this, the causal intervention $do(A_t := \texttt{Answer})$ is applied to simulate the alternative scenario. A key insight is that a definitive diagnosis simplifies to a practical test of the counterfactual answer's quality, sidestepping complex utility calculations. A correct counterfactual answer implies that the agent's knowledge state was indeed sufficient, whereas an incorrect answer implies it was not.

The $do(A_t := \texttt{Answer})$-operator is implemented via the answer intervention instruction. This instruction is designed to compel the agent to synthesize a final answer using only its current knowledge state $S_t$. The detailed prompt design is provided in Appendix~\ref{sec:prompt}.

\paragraph{\textbf{Diagnosing Under-search}}
In the case of an under-search error, the diagnosis is revealed by the factual outcome. When the agent's action is $A_t = \texttt{Answer}$ and the resulting answer is inaccurate, the latent state is immediately inferred to be $K_t = \text{Insufficient}$. No intervention is needed to identify the error itself.

Intervention is nonetheless critical for the subsequent task of remediation, which is to generate a corrected trajectory for the alignment dataset. This task addresses the counterfactual question of what the superior outcome would have been had the agent searched instead. The intervention $do(A_t := \texttt{Search})$ is therefore applied to generate this outcome. It is implemented via the search intervention instruction, which guides the agent to execute its higher-utility action. The resulting corrected, counterfactual trajectory then serves as the positive example for aligning the agent's decision policy. The detailed prompt design is provided in Appendix~\ref{sec:prompt}.

\subsection{Decision Boundary Alignment from Causal Feedback}
\label{sec:dpo_construction}

% \zxy{what are the Challenges/Motivations of the section (e.g., why the problem is important, why it is challenging, what are the shortcomings of existing solutions, why this module is needed, why do you use the provided technology, etc), otherwise your technical contributions look trivial.}
While causal diagnosis reveals where the agent’s decisions deviate from optimality, effective improvement requires aligning its policy with these causal signals to refine the decision boundary.
The causal interventions performed during diagnosis do more than identify errors; they generate counterfactual trajectories that provide a powerful signal for policy alignment. By comparing the agent's suboptimal factual path with the superior counterfactual path revealed through intervention, we obtain a clear causal preference. This section details the construction of a targeted preference dataset from these signals to align the agent's decision policy, $\pi$.

The dataset is composed of preference pairs, $(\text{preferred sequence } y_c$, $\text{rejected sequence } y_r)$, derived directly from the diagnosed decision boundary failures.

\paragraph{\textbf{Preferences from Over-search Errors}}
When an over-search error at state $S_t$ is diagnosed, the intervention $do(A_t := \texttt{Answer})$ reveals a more efficient path to a correct answer. For this case, a preference pair is formed to reward efficiency:
\begin{itemize}[leftmargin=*, itemsep=0pt, parsep=0pt, topsep=0pt]
    \item The \textbf{preferred sequence ($y_c$)} is the shorter, counterfactual trajectory generated by the intervention, which directly answers from state $S_t$.
    \item The \textbf{rejected sequence ($y_r$)} is the agent's original, factual trajectory, which includes the causally redundant \texttt{Search} step.
\end{itemize}

\paragraph{\textbf{Preference from Under-search Errors}}
When an under-search error is diagnosed, the intervention $do(A_t := \texttt{Search})$ is applied to generate a corrective, higher-utility trajectory. This forms a preference pair that rewards thoroughness:
\begin{itemize}[leftmargin=*, itemsep=0pt, parsep=0pt, topsep=0pt]
    \item The \textbf{preferred sequence ($y_c$)} is the superior, counterfactual trajectory resulting from the intervention, which includes the necessary search step.
    \item The \textbf{rejected sequence ($y_r$)} is the agent's original, factual trajectory that terminated prematurely with a low-quality answer.
\end{itemize}

The resulting preference dataset, $\mathcal{D} = \{(x, y_c, y_r)\}$, where $x$ represents the state history leading to the decision, is used to fine-tune the policy $\pi_{\theta}$ by minimizing the Direct Preference Optimization~\cite{rafailov2023direct} loss:
\begin{equation}
\label{eq:dpo_loss_alt}
\mathcal{L}_{\text{DPO}}(\pi_\theta; \pi_{\text{ref}}) = - \mathbb{E}_{(x, y_c, y_r) \sim \mathcal{D}} \left[ \log \sigma\left(\beta \log \frac{\pi_\theta(y_c|x)\pi_{\text{ref}}(y_r|x)}{\pi_\theta(y_r|x)\pi_{\text{ref}}(y_c|x)}\right) \right]
\end{equation}
where $\pi_{\text{ref}}$ is the original agent policy, $\beta$ is a temperature parameter, and $\sigma$ is the logistic function. This training process incentivizes the policy to adopt the preferred, causally optimal decision paths while penalizing the suboptimal ones, thereby directly refining its judgment at the decision boundary.

\section{Experiments}
\label{sec:experiments}

This section presents a comprehensive experimental evaluation conducted on three question-answering (QA) datasets. Our experiments are designed to answer the following research questions (RQs) systematically:
\begin{itemize}[leftmargin=*, itemsep=0pt, parsep=0pt, partopsep=0pt]
    \item \textbf{RQ1:} To what extent do decision boundary errors exist in current deep search agents?
    \item \textbf{RQ2:} Which question characteristics are more likely to cause decision boundary errors?
    \item \textbf{RQ3:} What is the impact of our proposed Decision Boundary Alignment on agent accuracy, efficiency, and decision quality?
    \item \textbf{RQ4:} What is the relationship and discrepancy between an agent's knowledge boundary and its decision boundary?
    \item \textbf{RQ5:} How does the number of reasoning steps affect the prevalence of decision boundary errors?
    % \item \textbf{RQ6:} What is the correlation between model uncertainty and the occurrence of decision boundary errors?
\end{itemize}
In the following subsections, we first detail our experimental setup, after which dedicated analyses are provided to address each of these research questions.

\subsection{Experimental Settings}
\subsubsection{\textbf{Dataset}}
% We conduct our experiments on three open-domain QA datasets that cover both single-hop and multi-hop reasoning scenarios. We use Natural Questions (NQ)~\cite{kwiatkowski2019natural} for single-hop evaluation, and both HotpotQA~\cite{yang2018hotpotqa} and 2WikiMultiHopQA~\cite{ho2020constructing} to assess multi-hop reasoning capabilities. 
Our experiments are conducted on three benchmark datasets selected to assess performance across varying levels of reasoning complexity. We use \textbf{Natural Questions (NQ)}~\cite{kwiatkowski2019natural} to evaluate single-hop retrieval performance. For multi-hop reasoning, we employ \textbf{HotpotQA}~\cite{yang2018hotpotqa} and \textbf{2WikiMultiHopQA}~\cite{ho2020constructing}, both of which require aggregating information from multiple documents. We adopt the standard settings for all datasets, utilizing the official training and development splits.
\subsubsection{\textbf{Baselines}}
Our evaluation is designed to first validate the existence of decision boundary errors and then to assess our proposed alignment method. To this end, we employ two distinct and representative workflows. Search-R1 embodies the approach of using end-to-end reinforcement learning (RL) to directly optimize a search agent. This workflow serves as our primary testbed, where we compare different models from the Qwen series~\cite{team2024qwen2} (Qwen2.5-7B/14B-base vs. -RL) and, crucially, evaluate our Decision Boundary Alignment. In contrast, Search-O1 represents the paradigm of integrating search tools into powerful general-purpose models. This allows us to analyze the behavior of leading closed-source models, including OpenAI's GPT-4o~\cite{hurst2024gpt}, GPT-4o-mini, Google's Gemini-2.5-Flash~\cite{comanici2025gemini}, and DeepSeek V3~\cite{liu2024deepseek}, in a search-augmented setting.

% \subsubsection{\textbf{Metrics}}
% \label{ssec:metrics}
% We evaluate agent performance using several key metrics. We measure accuracy with Exact Match (EM), and efficiency with total inference time and the Average Search Queries (ASQ). To specifically quantify decision boundary errors, we calculate the Over-Search Rate (OSR) and the Under-Search Rate (USR), respectively.

\subsubsection{\textbf{Metrics}}
\label{ssec:metrics}
Our evaluation protocol covers three critical aspects of agent performance. First, to assess the fundamental accuracy, we utilize the \textbf{Exact Match (EM)} metric. Second, considering the practical efficiency of agents, we evaluate efficiency using \textbf{Total Inference Time} and the \textbf{Average Search Queries (ASQ)}, aiming for high accuracy with minimal retrieval costs. Finally, to provide a granular analysis of the agent's decision boundary errors, we calculate the \textbf{Over-Search Rate (OSR)} and \textbf{Under-Search Rate (USR)}.

\subsubsection{\textbf{Implementation Details}}
Our retrieval pipeline is built upon a corpus from the Wikipedia dump (\texttt{Wikidump-20250901}~\cite{wikimedia_enwiki_2025}), which is segmented into sentence-level chunks. For retrieval, we employ the E5 embedding model~\cite{wang2023text} to fetch the top-3 most relevant chunks per query. Our baseline Search-R1 workflow follows the implementation of the open-source FlashRAG framework. To perform our Decision Boundary Alignment, we constructed a preference dataset of 20,000 examples by sampling 10,000 instances from each of the NQ and HotpotQA training sets. This preference data was generated using the RL-tuned Qwen2.5-7B-RL and Qwen2.5-14B-RL models, which were then further fine-tuned on this same dataset. We perform the alignment training via DPO for 3 epochs. Key hyperparameters include a DPO $\beta$ of 0.3, and a LoRA configuration \cite{qing2024alphalora} with rank $r$ of 64 and $\alpha$ of 128 for parameter-efficient tuning.

\begin{table*}[h]
\centering
\caption{Performance comparison of eight models across different deep search agents on the NQ, HotpotQA, and 2WikiMultiHopQA datasets. ASQ denotes Average Search Queries.}
\label{tab:model_performance}
\setlength{\tabcolsep}{4.5pt}
\resizebox{\textwidth}{!}{%
\begin{tabular}{@{}ll|ccccc|ccccc|ccccc@{}}
\toprule
\multirow{2}{*}{\textbf{Workflow}} & \multirow{2}{*}{\textbf{Model}} & \multicolumn{5}{c|}{\textbf{NQ}} & \multicolumn{5}{c|}{\textbf{HotpotQA}} & \multicolumn{5}{c}{\textbf{2WikiMultiHopQA}} \\
\cmidrule(lr){3-7} \cmidrule(lr){8-12} \cmidrule(lr){13-17}
& & \textbf{EM} & \textbf{Time} & \textbf{ASQ} & \textbf{OSR} & \textbf{USR} & \textbf{EM} & \textbf{Time} & \textbf{ASQ} & \textbf{OSR} & \textbf{USR} & \textbf{EM} & \textbf{Time} & \textbf{ASQ} & \textbf{OSR} & \textbf{USR} \\
\midrule
\multirow{4}{*}{Search-R1} & Qwen2.5-7b-base & 0.174 & 269.77 & 1.142 & 0.1265 & 0.7651 & 0.197 & 316.05 & 1.817 & 0.1712 & 0.7167 & 0.253 & 352.44 & 2.342 & 0.2721 & 0.5793 \\
& Qwen2.5-7b-RL & 0.381 & 464.02 & 2.014 & 0.3280 & 0.5210 & 0.407 & 638.12 & 2.640 & 0.3020 & 0.5430 & 0.457 & 644.81 & 3.200 & 0.3920 & 0.4090 \\
& Qwen2.5-14b-base & 0.260 & 727.55 & 1.714 & 0.1930 & 0.6590 & 0.292 & 834.74 & 2.168 & 0.2202 & 0.6206 & 0.345 & 816.80 & 2.553 & 0.2358 & 0.5435 \\
& Qwen2.5-14b-RL & 0.440 & 331.19 & 1.016 & 0.3233 & 0.5015 & 0.421 & 540.58 & 1.457 & 0.2928 & 0.5216 & 0.569 & 768.11 & 2.168 & 0.3195 & 0.3630 \\
\midrule
\multirow{4}{*}{Search-O1} & GPT-4o & 0.270 & 40.86 & 1.226 & 0.1473 & 0.6748 & 0.440 & 63.31 & 1.785 & 0.2592 & 0.4311 & 0.520 & 87.86 & 2.653 & 0.1581 & 0.4729 \\
& GPT-4o-mini & 0.170 & 325.65 & 1.230 & 0.1238 & 0.8314 & 0.360 & 134.97 & 1.772 & 0.1859 & 0.6407 &  0.290 & 130.83 & 2.308 & 0.1224 & 0.6805 \\
& Gemini-2.5-Flash & 0.350 & 198.44 & 0.591 & 0.1600 & 0.7198 & 0.390 & 281.86 & 0.404 & 0.1167 & 0.5415 & 0.390 & 210.69 & 0.390 & 0.2500 & 0.4952 \\
& deepseek-V3 & 0.250 & 126.51 & 1.141 & 0.2906 & 0.6289 & 0.470 & 195.28 & 1.383 & 0.4374 & 0.4122 & 0.460 & 180.48 & 1.595 & 0.0218 & 0.5263 \\
\bottomrule
\end{tabular}%
}
\end{table*}

\subsection{Decision Boundary Analysis (RQ1)}
\label{sec:analysis}
We conducted extensive experiments to investigate the decision boundary problem across various deep search agents and dataset settings to answer RQ1. Specifically, We evaluated two primary categories of state-of-the-art agents. The first is Search-R1, which employs outcome-based reinforcement learning to optimize search capabilities. For this category, we compare the Qwen2.5-7B and Qwen2.5-14B models before and after RL fine-tuning. The second category, Search-O1, includes powerful, closed-source models such as the GPT series, which serve as strong baseline agents. The comprehensive results are presented in Table~\ref{tab:model_performance}, from which we derive the following key conclusions.

\paragraph{\textbf{The Decision Boundary Problem is Pervasive}}
The results in Table~\ref{tab:model_performance} unequivocally demonstrate that flawed decision-making is a pervasive challenge. All tested agents, regardless of architecture, scale, or training paradigm, exhibit significant levels of both OSR and USR. For instance, even a powerful model like GPT-4o shows a USR of 0.670 on NQ, indicating a frequent failure to initiate a search when necessary. This establishes the decision boundary problem not as a niche issue but as a fundamental bottleneck for current search agents. These errors directly degrade performance: a high USR limits an agent's potential accuracy by failing to retrieve critical information, while a high OSR harms efficiency by wasting computational resources on unnecessary queries. 
% A detailed case study is provided in Appendix~\ref{sec:case}.

\paragraph{\textbf{Outcome-based RL Reveals a Critical Trade-off}}
We examined whether outcome-based RL mitigates decision boundary errors and found that it introduces a critical trade-off. On the NQ dataset, RL dramatically boosts the accuracy of Qwen2.5-7B (EM from 0.190 to 0.410), primarily by teaching the agent to be more thorough. This is reflected in a substantial reduction in USR (from 0.672 to 0.505) as the agent learns to search more often. However, this aggressive search policy comes at a steep efficiency cost: the OSR increases (from 0.196 to 0.263), and the average search queries (ASQ) rise by over 60\%. This finding reveals that relying on accuracy-centric rewards can correct the critical issue of under-search, but at the expense of pushing the decision boundary too far. By incentivizing correctness without penalizing cost, the model alleviates under-search only to exacerbate over-search.

\paragraph{\textbf{Universality Across Scales and Workflows}}
Our empirical analysis indicates that the challenge of optimal decision boundary placement is pervasive, persisting across varying model scales and distinct workflows. While scaling up model parameters (e.g., from Qwen2.5-7B to 14B) improves general capabilities, Table 1 shows that it does not resolve the conflict between efficiency and accuracy: larger models still exhibit significant rates of both over-search and under-search. Furthermore, this trade-off transcends specific pipelines. Whether employing the training-centric Search-R1 workflow or the inference-centric Search-O1 workflow (including GPT-4o, Gemini-2.5-Flash, and DeepSeek-V3), no agent achieves a balance where both errors are negligible. This universality suggests that the bottleneck is not merely model capability, but the fundamental difficulty of calibrating the optimal decision boundary.

% \paragraph{\textbf{Impact of Model Scale and Workflow}}
% Our experiments reveal that model scale helps to better calibrate this trade-off. After RL fine-tuning, the larger Qwen2.5-14B model not only surpasses its 7B counterpart in accuracy but also demonstrates far superior efficiency (ASQ of 1.02 vs. 2.01 on NQ). This suggests that larger models may possess a more nuanced internal representation of their own knowledge boundaries, enabling them to reduce under-search without a proportional increase in over-search. Furthermore, different workflows expose distinct, innate decision-making profiles. Gemini-2.5-Flash, for example, operates with an extremely low ASQ, indicating a strong preference for its parametric knowledge. This minimizes OSR but results in a high USR, creating the profile of a highly efficient but often ``confidently wrong'' agent. This diversity underscores that there is no one-size-fits-all decision boundary; it is highly dependent on a model's inherent characteristics.

\paragraph{\textbf{Influence of Task Complexity}}
Finally, comparing performance on NQ versus the multi-hop HotpotQA dataset highlights how task complexity exacerbates decision-making challenges. As expected, all agents increase their search frequency (higher ASQ) on the more complex HotpotQA task, showing a basic ability to adapt. However, this adaptation is imperfect. The persistently high OSR and USR values on HotpotQA indicate that as reasoning chains grow longer, an agent's ability to precisely determine the necessity of a search at each intermediate step diminishes significantly.

\begin{table}[!t]
\centering
\caption{Performance analysis across different data characteristics. We analyze the agent's decision boundary errors based on question difficulty, category, and the required number of supporting facts.}
\label{tab:data_characteristics}
\setlength{\tabcolsep}{4.5pt}
\resizebox{\columnwidth}{!}{%
\begin{tabular}{@{}llcccccc@{}}
\toprule
\textbf{Factor} & \textbf{Type} & \textbf{EM} & \textbf{F1} & \textbf{ACC} & \textbf{ASQ} & \textbf{OSR} & \textbf{USR} \\
\midrule
\multirow{3}{*}{\parbox{1.5cm}{\centering\textbf{Difficulty Level}}} & Easy & 0.500 & 0.623 & 0.523 & 2.318 & 0.402 & 0.402 \\
& Medium & 0.506 & 0.570 & 0.529 & 2.674 & 0.394 & 0.423 \\
& Hard & 0.453 & 0.552 & 0.480 & 2.598 & 0.402 & 0.469 \\
\midrule
\multirow{2}{*}{\parbox{1.5cm}{\centering\textbf{Question Category}}} & Comparison & 0.717 & 0.806 & 0.754 & 2.594 & 0.738 & 0.193 \\
& Bridge & 0.444 & 0.526 & 0.465 & 2.582 & 0.319 & 0.481 \\
\midrule
\multirow{4}{*}{\parbox{1.5cm}{\centering\textbf{\# Supporting Facts}}} & 2 & 0.520 & 0.604 & 0.546 & 2.526 & 0.423 & 0.399 \\
& 3 & 0.464 & 0.541 & 0.469 & 2.680 & 0.356 & 0.464 \\
& 4 & 0.360 & 0.445 & 0.413 & 2.867 & 0.293 & 0.560 \\
& $\ge$5 & 0.300 & 0.417 & 0.300 & 2.800 & 0.100 & 0.700 \\
\bottomrule
\end{tabular}%
}

\end{table}

\begin{table*}[h]
    \setlength{\tabcolsep}{4.5pt}
    \centering
    \caption{Main results and ablation studies for our decision boundary alignment method applied to Search-R1. For each metric, the best-performing result is in \textbf{bold} and the second-best is \underline{underlined}.}
    \label{tab:model_alignment_ablation_final_v4}
    \resizebox{\textwidth}{!}{%
    \begin{tabular}{@{}l|ccccc|ccccc|ccccc@{}}
    \toprule
    \multirow{2}{*}{\textbf{Model}} & \multicolumn{5}{c|}{\textbf{NQ}} & \multicolumn{5}{c|}{\textbf{HotpotQA}} & \multicolumn{5}{c}{\textbf{2WikiMultiHopQA}} \\
    \cmidrule(lr){2-6} \cmidrule(lr){7-11} \cmidrule(lr){12-16}
    & \textbf{EM} & \textbf{Time} & \textbf{ASQ} & \textbf{OSR} & \textbf{USR} & \textbf{EM} & \textbf{Time} & \textbf{ASQ} & \textbf{OSR} & \textbf{USR} & \textbf{EM} & \textbf{Time} & \textbf{ASQ} & \textbf{OSR} & \textbf{USR} \\
    \midrule
    Qwen2.5-7B-RL & 0.381 & 464.02 & 2.014 & 0.3280 & 0.5210 & 0.407 & 638.12 & 2.640 & 0.3020 & 0.5430 & 0.457 & 644.81 & 3.200 & 0.3920 & 0.4090 \\
    \textbf{+ DAS} & \textbf{0.394} & \underline{432.06} & \underline{1.896} & \underline{0.3080} & \underline{0.5180} & \textbf{0.414} & \underline{593.45} & \underline{2.568} & \underline{0.2790} & \underline{0.5370} & \textbf{0.478} & \underline{632.71} & \underline{3.113} & \underline{0.3746} & \underline{0.4030} \\
    \quad w/o under-search data & \underline{0.392} & \textbf{415.52} & \textbf{1.758} & \textbf{0.2883} & 0.5356 & \underline{0.412} & \textbf{578.11} & \textbf{2.411} & \textbf{0.2595} & 0.5591 & \underline{0.476} & \textbf{615.90} & \textbf{2.945} & \textbf{0.3528} & 0.4242 \\
    \quad w/o over-search data & 0.387 & 533.62 & 2.215 & 0.3580 & \textbf{0.4410} & 0.409 & 733.84 & 2.855 & 0.3320 & \textbf{0.4630} & 0.461 & 741.53 & 3.410 & 0.4220 & \textbf{0.3290} \\
    \cmidrule(lr){1-16}
    Qwen2.5-14B-RL & 0.440 & 331.19 & 1.016 & 0.3233 & 0.5015 & 0.421 & 540.58 & 1.457 & 0.2928 & 0.5216 & 0.569 & 768.11 & 2.168 & 0.3195 & 0.3630 \\
    \textbf{+ DAS} & \textbf{0.454} & \underline{308.02} & \underline{0.955} & \underline{0.3039} & \underline{0.4965} & \textbf{0.428} & \underline{502.74} & \underline{1.413} & \underline{0.2723} & \underline{0.5164} & \textbf{0.580} & \underline{714.34} & \underline{2.103} & \underline{0.2971} & \underline{0.3594} \\
    \quad w/o under-search data & \underline{0.452} & \textbf{295.18} & \textbf{0.861} & \textbf{0.2854} & 0.5147 & \underline{0.426} & \textbf{489.25} & \textbf{1.302} & \textbf{0.2541} & 0.5388 & \underline{0.578} & \textbf{698.55} & \textbf{1.986} & \textbf{0.2798} & 0.3776 \\
    \quad w/o over-search data & 0.446 & 380.87 & 1.251 & 0.3533 & \textbf{0.4781} & 0.423 & 621.67 & 1.682 & 0.3228 & \textbf{0.4952} & 0.572 & 883.33 & 2.455 & 0.3495 & \textbf{0.2830} \\
    \bottomrule
    \end{tabular}%
    }
\end{table*}

\subsection{Impact of Data Characteristics on Decision Errors (RQ2)}
\label{sec:data_characteristics_analysis}

We conduct extensive experiments to investigate the decision boundary errors on different type of datasets to answer RQ2. Specifically, We perform a fine-grained analysis of the agent's performance on subsets of the HotpotQA dataset, partitioned along three distinct dimensions of complexity. This allows us to create controlled testbeds to isolate specific failure modes.

\begin{itemize}[leftmargin=*, itemsep=0pt, parsep=0pt, partopsep=0pt]
    \item \textbf{Difficulty Level:} This dimension categorizes questions as \texttt{Easy}, \texttt{Medium}, or \texttt{Hard}, based on human annotations established during the dataset's creation. The level reflects the human-perceived difficulty of the search task and generally correlates with the number of search actions required.
    \item \textbf{Question Category:} This dimension distinguishes between two reasoning types. \texttt{Comparison} questions require retrieving multiple, independent pieces of information. In contrast, \texttt{Bridge} questions necessitate a sequential reasoning path where the queries are dependent on one another.
    \item \textbf{\# Supporting Facts:} This metric refers to the number of distinct text paragraphs required to derive the final answer. It directly correlates with the searching burden and the minimum number of retrieval actions needed to complete the task.
\end{itemize}

The experimental results are summarized in Table~\ref{tab:data_characteristics}, and our key observations are detailed below.

% \paragraph{Impact of Question Difficulty}
% As the general difficulty increases from Easy to Hard, the agent's failure mode shifts decisively. While accuracy metrics (EM, F1, ACC) decline slightly, the most significant change is the steady increase in the USR from 0.402 to 0.469. The OSR and Average Search Queries (ASQ) remain relatively stable. This suggests that as questions become implicitly harder, the agent's primary weakness is its failure to recognize the need for more rigorous information gathering. It is prone to halting the search process prematurely and attempting to answer with insufficient evidence.
As the general difficulty increases from Easy to Hard, the agent's failure mode shifts. While accuracy metrics (EM, F1, ACC) decline slightly, the most significant change is the steady increase in the USR from 0.402 to 0.469. The OSR and the average search queries (ASQ) remain relatively stable. This suggests that as questions become implicitly harder, the agent's primary weakness is its failure to recognize the need for more rigorous information gathering. It is prone to halting the search process prematurely and attempting to answer with insufficient evidence.

\paragraph{\textbf{Impact of Question Category}}
\texttt{Comparison} and \texttt{Bridge} questions exhibit two starkly different decision-making pathologies. \texttt{Comparison} questions, which involve parallel information retrieval, yield high accuracy (0.754 ACC) but suffer from an extremely high \textbf{OSR} of 0.738. This indicates that while the agent correctly identifies the need for multiple pieces of information, its decision process is highly inefficient, leading to redundant searches. Conversely, \texttt{Bridge} questions, which require a dependent reasoning path, show an inverse error pattern. The OSR is low (0.319), but the \textbf{USR} is alarmingly high at 0.481. This reveals a critical failure in multi-step reasoning: the agent struggles to use the result of one search to inform the necessity of the next. Its primary failure is not inefficient searching, but rather \textit{not knowing it needs to continue searching at all}.

\paragraph{\textbf{Impact of the Number of Supporting Facts}}
This dimension directly measures the information-gathering burden. As the required number of supporting facts increases from 2 to 5 or more, a clear pattern of insufficient persistence emerges. Although the agent slightly increases its search attempts (ASQ rises from 2.526 to 2.800), this is inadequate for the task's escalating complexity. The result is a dramatic trade-off: OSR plummets from 0.423 to just 0.100, while the USR skyrockets from 0.399 to a critical 0.700. The low OSR for questions needing $\ge$5 facts is not a sign of efficiency; it is a symptom of the agent \textit{giving up too early}. It executes initial searches but lacks the decision-making logic to persist until all necessary information is collected, highlighting a decision boundary biased towards premature termination on information-intensive questions.

\subsection{Decision Boundary Alignment (RQ3)}
\label{ssec:alignment_results}

This section evaluates the efficacy of our proposed decision boundary alignment method to answer RQ3. Specifically, we apply our alignment technique to two deep search agents: Qwen2.5-7B-RL and Qwen2.5-14B-RL, both trained using Search-R1. To dissect the contribution of each component, we conduct an ablation study with the following variants:

\begin{itemize}[leftmargin=*, itemsep=0pt, parsep=0pt, partopsep=0pt]
\item \textbf{DAS}: Our proposed Decision Boundary Alignment method, which aligns the model using preference data that targets both under-search and over-search errors.
\item \textbf{w/o under-search data}: A variant that omits under-search data, relying solely on over-search preference data for alignment.
\item \textbf{w/o over-search data}: A variant that omits over-search data, relying solely on under-search preference data for alignment.
\end{itemize}

The experimental results are summarized in Table~\ref{tab:model_alignment_ablation_final_v4} and the conclusion are listed as follows.

\paragraph{\textbf{Main Alignment Results}}
As demonstrated in Table~\ref{tab:model_alignment_ablation_final_v4}, our proposed DAS method yields substantial and consistent improvements. This performance gain is robust across both the Qwen2.5-7B-RL and Qwen2.5-14B-RL models and holds for all three QA datasets. Specifically, DAS achieves the highest accuracy, as measured by exact match (EM), in nearly all configurations. Furthermore, the method enhances operational efficiency by significantly reducing both inference time and the average number of search queries (ASQ) compared to the baseline RL models. Crucially, DAS effectively mitigates decision boundary errors by simultaneously lowering the over-searching rate (OSR) and the under-searching rate (USR), resulting in a more calibrated and reliable search agent.

\paragraph{\textbf{Ablation Study}}
Our ablation study dissects the contribution of each data component, revealing a critical trade-off. Removing the under-search preference data (\textit{w/o under-search data}) compels the agent to adopt a highly aggressive and efficient policy, achieving the best scores on all efficiency-related metrics (time, ASQ, and OSR). Conversely, omitting the over-search data (\textit{w/o over-search data}) induces an overly cautious agent. This variant exhibits a stronger inclination to search, minimizing the USR but at the expense of overall efficiency. These two signals function antagonistically; emphasizing one degrades performance in the other's domain. Therefore, only the complete DAS framework, which concurrently balances both error types, achieves a superior synthesis of top-tier accuracy and well-rounded efficiency.

\begin{table}[!t]
\centering
\caption{Correlation between search decisions and knowledge boundary. \textbf{Acc} measures the agent's accuracy when forced to answer without searching (Probe 1). \textbf{Confidence} measures the frequency (\% 'Yes') the agent claims to know the answer (Probe 2).}
\label{tab:knowledge_vs_decision}
\begin{tabular}{@{}l c c c@{}}
\toprule
\textbf{Decision Category} & \textbf{Proportion} & \textbf{ACC} & \textbf{Confidence} \\
\midrule
% Over-search & 19.02\% & 0.5069 & 0.5376 \\
% Correct-search & 80.98\% & 0.4198 & 0.4724 \\
Over-search & 19.02\% & 0.2356 & 0.5750 \\
Effective search & 80.98\% & 0.1423 & 0.4859 \\
\bottomrule
\end{tabular}
\end{table}

\subsection{Analysis of the Knowledge-Decision Gap (RQ4)}
\label{ssec:knowledge}
The relationship between an agent's knowledge boundary and its decision boundary remains an open question. To investigate the gap between them(RQ4), we categorize agent searches into two types: Over-search (unnecessary) and Effective search (beneficial). We then probe the agent's knowledge using two metrics: \textit{ACC}, its accuracy when intervened to answer without searching, and \textit{Confidence}, its self-assessed ability to answer ('Yes/No'). The full prompts used in our study are available in Appendix~\ref{sec:prompt}.

Table~\ref{tab:knowledge_vs_decision} presents our findings. The results reveal a nuanced picture. We observed that Over-search, which constitutes 19.02\% of all actions, paradoxically shows higher agent ACC (0.2356 vs. 0.1423) and Confidence (0.5750 vs. 0.4859) compared to Effective search. This leads to a two-part conclusion. First, a correlation exists, as the agent's knowledge state clearly influences its search decisions. However, the low absolute ACC highlights a severe miscalibration: the agent still fails to answer correctly in over 76\% of these "unnecessary" search cases. We conclude that the Knowledge-Decision Gap stems from the agent's poor self-awareness, which remains a key bottleneck for current architectures.

\begin{figure}[!t]
    \centering
    \includegraphics[width=0.98\linewidth]{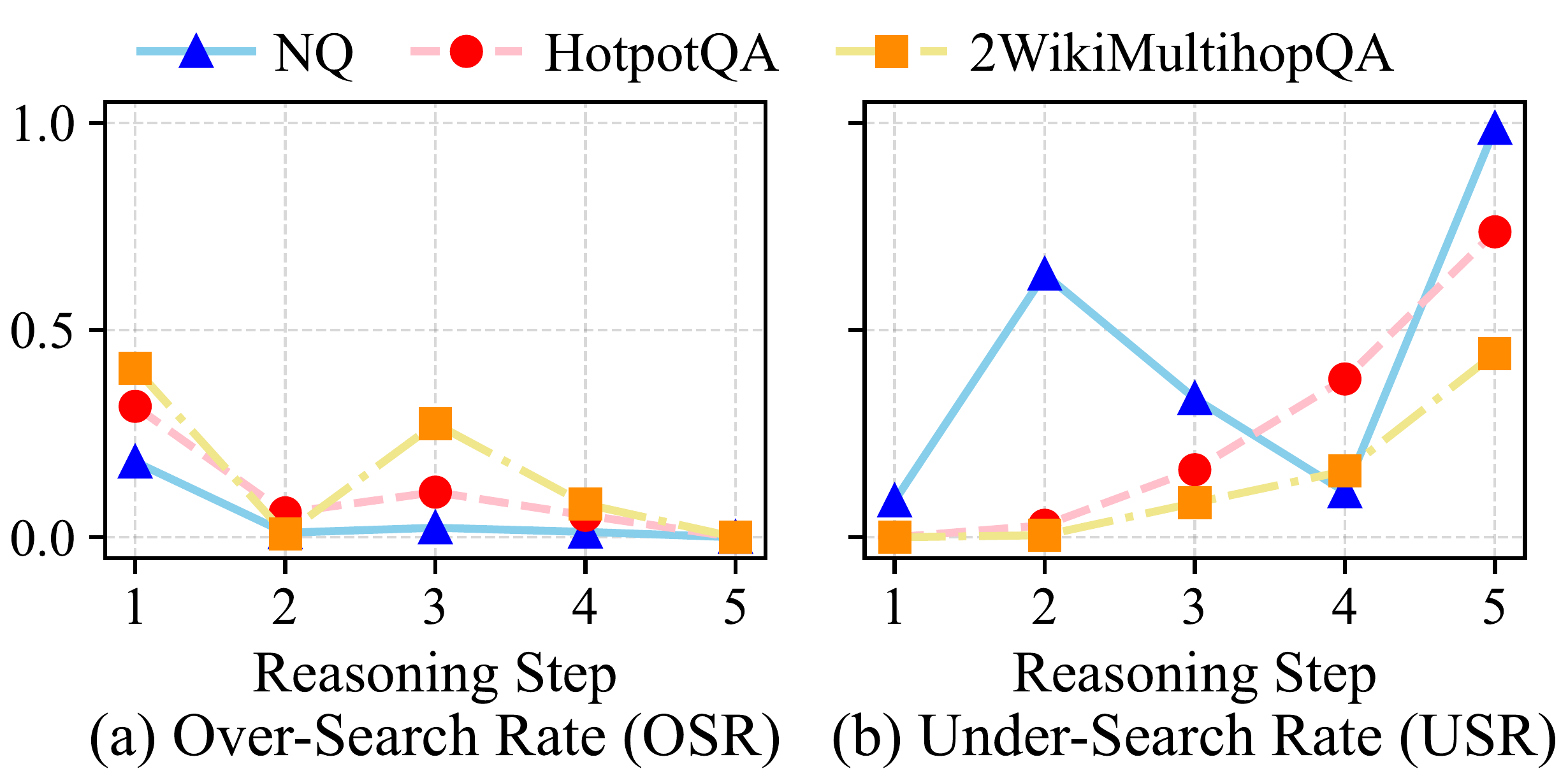}
    \caption{Distribution of decision boundary errors (OSR and USR) across reasoning steps on NQ, HotpotQA, and 2WikiMultihopQA (2Wiki) datasets.}
    \label{fig:step_analysis}
\end{figure}

\subsection{Step-wise Analysis of Decision Errors (RQ5)}
\label{sec:step_wise_analysis}

To dissect the dynamics of the \textbf{decision boundary problem} during an agent's autonomous reasoning process, we performed a step-wise analysis. This approach allows us to observe how the two core facets of this problem, \textbf{OSR} and \textbf{USR}, are distributed at each step. By comparing single-hop and multi-hop tasks, we aim to reveal how the dominant failure mode of the decision boundary evolves as the reasoning chain lengthens and task complexity increases. The results, presented in Figure~\ref{fig:step_analysis}, reveal distinct and evolving patterns.
% The results, presented in Table~\ref{tab:step_wise_errors_updated}, reveal distinct and evolving patterns.

\paragraph{\textbf{Over-search is Concentrated in the Initial Step.}}
A universal phenomenon across all datasets is that the OSR is highest at the first reasoning step. This indicates a general failure mode where agents reflexively initiate a search, even for questions that could potentially be answered directly from their parametric knowledge. While the OSR trends downwards significantly in subsequent steps, complex multi-hop tasks like 2WikiMultihopQA can exhibit a mid-process resurgence of over-search (0.274 at Step 3). This suggests that even when an agent correctly decides to continue its search, its mid-process decisions can still be inefficient and confused.

\paragraph{\textbf{Under-search Escalates with Reasoning Steps.}}
In contrast to OSR, the Under-search Rate (USR) systematically increases with the number of reasoning steps across all datasets. This indicates that agents are progressively more likely to answer incorrectly as task complexity increases, consistently misjudging information sufficiency for more difficult problems.

\paragraph{\textbf{Divergent Error Trajectories for Single-hop vs. Multi-hop Tasks.}}
The error patterns diverge significantly between task types. The single-hop NQ dataset shows an abrupt transition, where the error mode flips from a high OSR at step one to a critically high USR immediately after. In contrast, multi-hop datasets exhibit a gradual deterioration, with the USR climbing steadily at each subsequent step, indicating a progressive failure to assess information completeness.

% \paragraph{Under-search Escalates with Reasoning Steps.}
% In stark contrast to OSR, the USR demonstrates a clear and systematic upward trend as the number of reasoning steps increases. This trend holds true for all tested datasets. Since the number of steps required to stop and answer correlates with the intrinsic difficulty of a question, this escalating USR reveals a fundamental flaw: the more complex a problem is (requiring more reasoning steps), the higher the probability that the agent's final answer will be incorrect. The agent consistently fails to judge when it has gathered sufficient information for these harder problems.

% \paragraph{Divergent Error Trajectories for Single-hop vs. Multi-hop Tasks.}
% The pattern of these two error trends differs significantly in single-hop and multi-hop tasks. For the single-hop NQ dataset, the decision error pattern shows an abrupt transition: after the initial OSR peak, the error mode immediately flips to a critically high USR at Step 2. This represents a ``tipping point'' where the agent's judgment fails catastrophically after a single search. In contrast, the multi-hop datasets (HotpotQA, 2WikiMultihopQA) exhibit a \textbf{gradual deterioration}. The USR climbs steadily with each additional step, indicating a progressive, rather than sudden, failure to assess information completeness as the reasoning chain lengthens.

\section{Related Work}
\label{sec:related_work}

\subsection{Knowledge Boundaries in Agents}
\label{ssec:related_kb}
% Research on the \textit{Knowledge Boundary} of LLMs aims to delineate their reliable parametric knowledge to mitigate hallucinations~\cite{ni2024llms, li2024knowledge, yin2024benchmarking}. Boundary detection is often approached by quantifying model uncertainty or probing for model self-knowledge~\cite{kadavath2022language, qiu2024semantic, yin2023large}. RAG mitigates this boundary problem by augmenting the generation context with retrieved documents, extending the knowledge available during inference~\cite{lewis2020retrieval}. However, the conventional RAG paradigm relies on a static, non-adaptive retrieval policy, making it inefficient for complex tasks~\cite{asai2024self}. This limitation is a critical bottleneck for deep search agents, which require autonomous, multi-step interaction with search engines~\cite{li2025search, jin2025search}. For such agents, the problem evolves into a sequential decision-making challenge. The research question shifts from the static \textit{Knowledge Boundary} (i.e., whether a model \textit{can} answer) to a dynamic \textit{Decision Boundary} (i.e., whether an agent \textit{should} continue to search). Optimizing this decision policy is the core challenge we address.
Research on the \textit{Knowledge Boundary} of LLMs aims to delineate the scope of their reliable parametric knowledge to mitigate hallucinations~\cite{ni2024llms, li2024knowledge, yin2024benchmarking}. Boundary detection is often approached by quantifying model uncertainty via internal metrics like token entropy~\cite{kadavath2022language, qiu2024semantic} or by probing for model self-knowledge~\cite{yin2023large}. To mitigate this boundary problem, RAG method augments the context with documents retrieved from external corpora, extending the knowledge available during the inference stage~\cite{lewis2020retrieval}. However, the conventional RAG paradigm relies on a static retrieval mechanism, making it inefficient for complex tasks that demand iterative reasoning rather than simple one-off information fetching~\cite{asai2024self}. This limitation becomes a critical bottleneck for deep search agents, which require autonomous, multi-step interaction with web search engines~\cite{li2025search, jin2025search}. For such agents, the problem evolves from knowledge retrieval to a sequential decision-making challenge. The central research question thus shifts from the static \textit{Knowledge Boundary} (i.e., whether a model \textit{can} answer) to a dynamic \textit{Decision Boundary} (i.e., whether an agent \textit{should} continue to search). Optimizing this decision policy is the core challenge we address.

\subsection{Deep Search Agents}
Deep search agents are autonomous systems for addressing complex questions through multi-step information exploration and synthesis~\cite{huang2025deep, li2025towards, xi2025survey}. Current research aims to enhance their dynamic decision-making and query generation~\cite{zhang2025evolvesearch, sun2025zerosearch}. For instance, Search-o1~\cite{li2025search} employs structured iterative workflows to guide the agent's reasoning process. Other works like Search-R1~\cite{jin2025search}, R1-Searcher~\cite{song2025r1}, and ReasonRAG~\cite{zhang2025process} train the end-to-end reasoning policy using final answer accuracy as a reward signal. InForage~\cite{qian2025scent} and OTC~\cite{wang2025acting} introduce heuristics like penalizing numerous search calls to curb inefficiency.
However, prior work's outcome-centric focus treats the reasoning process as a black box, overlooking the optimality of each intermediate decision. 
% Relying solely on sparse rewards from the final answer, such methods fail to assign proper credit to specific retrieval steps, making it difficult to differentiate between necessary exploration and redundant querying. This lack of fine-grained feedback leads to aimless information accumulation, 
Relying on sparse final rewards, these approaches fail to distinguish essential exploration from redundant querying. This lack of granular supervision risks aimless information accumulation,
leaving the critical problem of search termination unaddressed. Our work confronts this by formalizing it as the decision boundary challenge, learning an explicit policy to determine when information is sufficient.
% However, the exclusive focus on outcome-based optimization in prior work treats the reasoning process as a black box, overlooking the optimality of each intermediate decision. This leaves the critical problem of when to terminate the search unaddressed. Our work directly confronts this issue by formalizing it as the decision boundary challenge, learning an explicit policy to determine when sufficient information has been gathered.
\section{Conclusion}
\label{sec:conclusion}
% In this paper, we formally define and address the \textit{Decision Boundary} problem in deep search agents, a critical challenge causing failures of \textit{over-search} (inefficiency) and \textit{under-search} (inaccuracy). We introduce a novel diagnostic methodology rooted in causal intervention. This approach retrospectively infers an agent's unobservable latent knowledge state by comparing its factual trajectory against a superior counterfactual one. Building on this causal diagnosis, we propose the \textbf{D}ecision Boundary \textbf{A}lignment (DAS) method, which calibrates an agent's judgment by constructing a preference dataset from these factual and counterfactual pairs. Our experiments provide the first quantitative evidence that these decision errors are a pervasive bottleneck in advanced agents. Crucially, the results demonstrate that DAS effectively mitigates both error types, achieving significant simultaneous gains in final-answer accuracy and overall efficiency. This research offers a foundational solution for creating more robust and economically viable autonomous agents for the web.
In this paper, we formally define and address the \textit{Decision Boundary} problem in deep search agents, a critical challenge causing failures of \textit{over-search} (inefficiency) and \textit{under-search} (inaccuracy). We introduce a novel diagnostic method rooted in causal intervention. This approach retrospectively infers an agent's latent knowledge state by comparing its factual trajectory against a superior counterfactual one. Building on this causal diagnosis, we propose the \textbf{D}ecision Boundary \textbf{A}lignment (DAS) method, which calibrates an agent's judgment by constructing a preference dataset from these factual and counterfactual pairs. Our experiments provide the first quantitative evidence that these decision errors are a pervasive bottleneck in advanced agents. Crucially, the results demonstrate that DAS effectively mitigates both error types, achieving significant simultaneous gains in final-answer accuracy and overall efficiency. This research offers a foundational solution for creating more robust and economically viable autonomous agents for the web.

\section*{Acknowledgements}
This research was partially supported by National Natural Science Foundation of China (No.62502404), Hong Kong Research Grants Council (Research Impact Fund No.R1015-23, Collaborative Research Fund No.C1043-24GF, General Research Fund No.11218325), Institute of Digital Medicine of City University of Hong Kong (No.9229503), and Huawei (Huawei Innovation Research Program). 

\clearpage

\bibliographystyle{ACM-Reference-Format}
\bibliography{9Reference}

% \clearpage

\appendix

\section{Uncertainty and the Decision Boundary}
\label{sec:uncertainty_analysis}
\begin{figure}[!t]
    \centering
    \includegraphics[width=0.8\columnwidth]{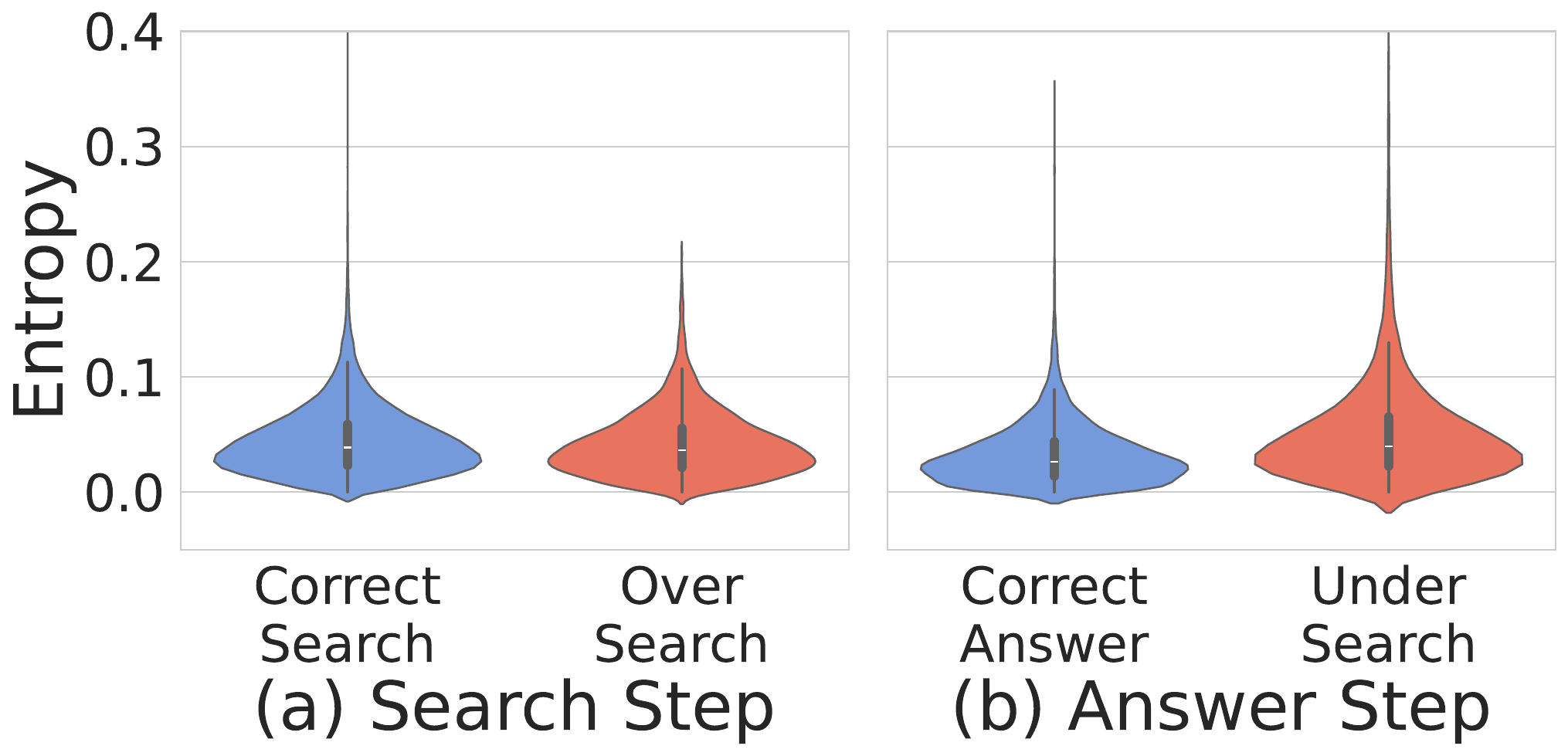}
    \vspace{-1em}
    \caption{Entropy distributions for correct vs. incorrect decisions at the Search Step (left) and Answer Step (right).}
    \vspace{-1em}
    \label{fig:entropy_violin}
\end{figure}

In this section, we investigate the relationship between the agent's output entropy and its decision boundary errors. We observe the distribution of entropy for correct versus incorrect decisions (over-search and under-search) during both the search and answer steps. The results, shown in Figure~\ref{fig:entropy_violin}, reveal the difference in these distributions. As illustrated, for the \textit{Search Step}, the entropy distributions for effective search and over-search are highly similar, indicating little difference. In contrast, for the \textit{Answer Step}, the overall entropy for correct-answer is lower than that for under-search, which suggests the model is more confident when answering correctly.

\begin{figure}[!htbp]
    \centering
    \includegraphics[width=0.8\columnwidth]{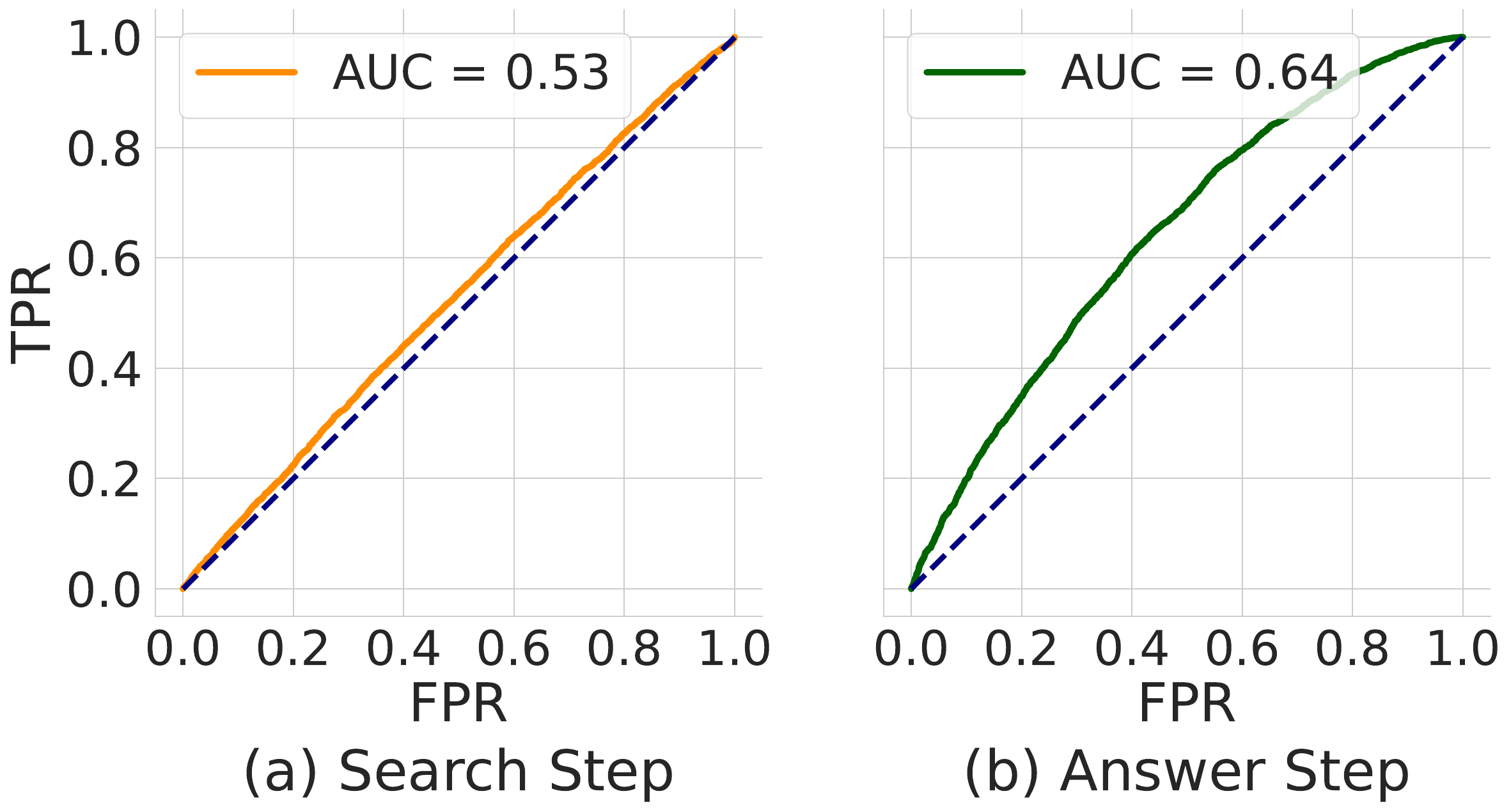}
    \vspace{-1em}
    \caption{ROC curves illustrating the effectiveness of using entropy to classify correct vs. incorrect decisions for the Search Step (a) and Answer Step (b).}
    \vspace{-1em}
    \label{fig:entropy_roc}
\end{figure}

Furthermore, we examined the potential of using entropy to mitigate decision boundary errors. We treated this as a classification task, using entropy as the feature to distinguish correct from incorrect decisions. The results are presented as ROC curves in Figure~\ref{fig:entropy_roc}. As shown, entropy offers almost no contribution to resolving the decision boundary problem for the \textit{Search Step} (AUC = 0.53). However, it provides a discernible, albeit moderate, benefit for the \textit{Answer Step} (AUC = 0.64), indicating its utility in identifying potential under-search errors.

\section{Prompt Instructions}
\label{sec:prompt}
We list the prompt utilized in our experiments as follows:
\begin{itemize}[leftmargin=*, itemsep=0pt, parsep=0pt, partopsep=0pt]
    \item Intervention Instruction. Building upon the probing methodology from Section~\ref{sec:probes}, we detail the specific instructions used to control the agent's actions. We present two interventions: the \textit{Answer Intervention Instruction} (Figure~\ref{fig:forced_answer_prompt}) and the \textit{Search Intervention Instruction} (Figure~\ref{fig:forced_search_prompt}).
    % The \textit{Forced-Search Probe} for Search-O1 is illustrated in Figure~\ref{fig:prompt_search_o1}.
    \item Knowledge Boundary. Building upon our discussion of the knowledge boundary in Section~\ref{ssec:knowledge}, we now present the specific prompts designed to evaluate an agent's knowledge recall and its meta-knowledge. The \textit{knowledge recall prompt} is shown in Figure~\ref{fig:prompt_knowledge_recall}, and the \textit{knowledge meta prompt} is displayed in Figure~\ref{fig:prompt_knowledge_meta}.

    % and the fact check prompt is shown in Figure~\ref{fig:prompt_llm_judge}.
\end{itemize}

% \begin{figure*}[!t]
% \centering
% \begin{AIbox}{Search Intervention Instruction}
% You are an advanced reasoning agent. Your task is to answer the given question by searching for information. 

% Follow these steps meticulously:
% \begin{enumerate}
%     \item \textbf{Initial Plan:} In your first \texttt{<think>} block \texttt{</think>}, you must decompose the user's question and formulate a clear, step-by-step plan.
    
%     \item \textbf{Critique \& Justify:} After receiving new information, you must \textbf{critically evaluate your current knowledge state} inside \texttt{<think>} and \texttt{</think>}. You must explicitly answer these questions in your thoughts:
%     \begin{itemize}
%         \item What is the specific knowledge gap I need to fill right now?
%         \item Is my previous reasoning free of errors or hallucinations?
%         \item Why is another search necessary to solve the problem?
%     \end{itemize}
    
%     \item \textbf{Targeted Search:} Based on your critique and justification, you must formulate a \textbf{specific, targeted, and non-repetitive} search query inside \texttt{<search>} and \texttt{</search>} to resolve the identified knowledge gap.
% \end{enumerate}

% Do not provide the answer yet.

% \noindent Question: \{question\}
% \end{AIbox}
% \caption{The system prompt for the \textit{Search Intervention Instruction}. This prompt guides the agent to re-evaluate its knowledge state, identify a specific gap, and formulate a necessary search query instead of answering prematurely.}
% \label{fig:forced_search_prompt}
% \end{figure*}
\begin{figure*}[!t]
\centering
\begin{AIbox}{Search Intervention Instruction}
\small
You are an advanced reasoning agent tasked with answering a question by searching for information. Follow these steps:
% The 'nosep' option removes vertical spacing in the list
\begin{enumerate}[nosep, leftmargin=*]
    \item \textbf{Initial Plan:} In your first \texttt{<think>} block, decompose the question and formulate a clear, step-by-step plan.
    
    \item \textbf{Critique \& Justify:} After receiving new information, critically evaluate your current knowledge state inside \texttt{<think>}. Explicitly answer:
    % The nested list is also made compact
    \begin{itemize}
        \item What is the specific knowledge gap I need to fill right now?
        \item Is my previous reasoning free of errors or hallucinations?
        \item Why is another search necessary to solve the problem?
    \end{itemize}
    
    \item \textbf{Targeted Search:} Based on your critique, formulate a specific, targeted, and non-repetitive search query inside \texttt{<search>} to resolve the identified gap.
\end{enumerate}
Do not provide the answer yet.
\noindent Question: \{question\}
\end{AIbox}
\caption{The system prompt for the \textit{Search Intervention Instruction}. This prompt guides the agent to re-evaluate its knowledge state, identify a specific gap, and formulate a necessary search query instead of answering prematurely.}
\label{fig:forced_search_prompt}
\end{figure*}

\begin{figure*}[!t]
\centering
\begin{AIbox}{Answer Intervention Instruction}
Answer the given question. 

You must conduct reasoning inside \texttt{<think>} and \texttt{</think>} first every time you get new information. 

After reasoning, since you have all the information needed, you must directly provide the answer inside \texttt{<answer>} and \texttt{</answer>}, without detailed illustrations or further searches. For example, \texttt{<answer>Beijing</answer>}.
% \vspace{1em} % Adds a bit of vertical space for separation
Question: \{question\}
\end{AIbox}
\caption{The system prompt for the \textit{Answer Intervention Instruction}. This prompt instructs the agent to bypass the search step and synthesize a final answer using only the knowledge contained within its current state.}
\label{fig:forced_answer_prompt}
\end{figure*}

\begin{figure*}[!t]
\centering
\begin{AIbox}{Knowledge Recall Prompt}
You are acting as a search engine. The following text is a user's raw request, which contains their intended search query inside \texttt{<search>} tags. Your task is to directly answer the user's intended query based on your internal knowledge. Provide a concise, direct answer. Do not explain your reasoning or mention that you are an AI.

% \vspace{1em}
% \noindent\hrulefill

\noindent\textbf{User's Full Request:} \texttt{\{question\}}

% \noindent\hrulefill
% \vspace{0.5em}

\noindent\textbf{Direct Answer:}
\end{AIbox}
\caption{The \textit{Knowledge Recall Prompt}. This prompt tests the model's ability to directly retrieve and state factual information from its internal knowledge base, mimicking the behavior of a search engine's direct answer feature.}
\label{fig:prompt_knowledge_recall}
\end{figure*}

\begin{figure*}[t]
\centering
\begin{AIbox}{Knowledge Meta-Cognition Prompt}
You are a helpful assistant. You will see a user's full request which contains a search query inside \texttt{<search>} tags. Based on the \textit{intended query} within that request, do you believe you have sufficient internal knowledge to answer it accurately without searching? Answer \textit{only} with 'Yes' or 'No'.

% \vspace{1em}
% \noindent\hrulefill

\noindent\textbf{User's Full Request:} \texttt{\{question\}}

% \noindent\hrulefill
% \vspace{0.5em}

\noindent\textbf{Answer:}
\end{AIbox}
\caption{The \textit{Knowledge Meta-Cognition Prompt}. This prompt is designed to probe the model's ability to self-assess the sufficiency of its own knowledge regarding a specific query, forcing a binary 'Yes' or 'No' response.}
\label{fig:prompt_knowledge_meta}
\end{figure*}

\clearpage

\end{document}